\documentclass[letterpaper,twocolumn,english,pre]{revtex4}
\usepackage[T1]{fontenc}
\usepackage[latin9]{inputenc}
\usepackage{xcolor}
\usepackage{pdfcolmk}
\usepackage{amstext}
\usepackage{graphicx}
\usepackage{amssymb}
\PassOptionsToPackage{normalem}{ulem}
\usepackage{ulem}

\makeatletter

\pdfpageheight\paperheight
\pdfpagewidth\paperwidth

\providecolor{lyxadded}{rgb}{0,0,1}
\providecolor{lyxdeleted}{rgb}{1,0,0}

\@ifundefined{textcolor}{}
{%
 \definecolor{BLACK}{gray}{0}
 \definecolor{WHITE}{gray}{1}
 \definecolor{RED}{rgb}{1,0,0}
 \definecolor{GREEN}{rgb}{0,1,0}
 \definecolor{BLUE}{rgb}{0,0,1}
 \definecolor{CYAN}{cmyk}{1,0,0,0}
 \definecolor{MAGENTA}{cmyk}{0,1,0,0}
 \definecolor{YELLOW}{cmyk}{0,0,1,0}
 }

\makeatother

\usepackage{babel}

\begin{document}

\title{Variational ansatz for quasispecies in the Eigen model}

\author{Min-Feng Tu}

\affiliation{Department of Physics, National Tsing Hua University, Hsinchu 30043,
Taiwan}

\author{Ching-I Huang}

\affiliation{Department of Physics, National Tsing Hua University, Hsinchu 30043,
Taiwan}

\author{Hsiu-Hau Lin}

\email{hsiuhau@phys.nthu.edu.tw}

\affiliation{Department of Physics, National Tsing Hua University, Hsinchu 30043,
Taiwan}

\author{Chun-Chung Chen}

\email{cjj@u.washington.edu}

\affiliation{Physics Division, National Center for Theoretical Sciences, Hsinchu
300, Taiwan}

\date{\today}
\begin{abstract}
We investigate the error threshold for the emergence of quasispecies
in the Eigen model. By mapping to an effective Hamiltonian ruled by
the {}``imaginary-time'' Schrödinger equation, a variational ansatz
is proposed and applied to calculate various properties associated
with the quasispecies. The variational ansatz gives correct prediction
for the survival population of the wild-type sequence and also reveals
an unexpected universal scaling behavior near the error threshold.
We compare the results from the variational ansatz with that from
numerical methods and find excellent agreement. Though the emergence
of the scaling behavior is not yet fully understood, it is remarkable
that the universal scaling function reigns even for relatively short
genome length such as $L=16$. Further investigations may reveal the
mechanism of the universal scaling and extract the essential ingredients
for the emergence of the quasispecies in molecular evolution. 
\end{abstract}
\maketitle

\section{Introduction}

Molecular perspective for biological evolution\cite{Crow70,Eigen71,Leuthausser87,Dominigo99,Dominigo01}
stired up interesting ideas and attracted much attention in recent
decades. In 1971, Eigen proposed a simple but profound model to study
the process of molecular replication and introduced the concept of
quasispecies\cite{Eigen71} which turned out to be crucial in understanding
the fundamental behavior of evolution. A quasispecies consists of
a wild type with large fitness accompanied by a large number of mutant
types in sequence space\cite{Nowak92}. The Eigen model refines our
understanding of the evolutionary dynamics: Selection moves the population
towards the better adapted mutants and explores for even better ones
in the sequence space by random mutations. The Eigen model is particularly
suitable for viral evolution\cite{Dominigo99,Dominigo01} and population
genetics\cite{Crow70,Hofbauer98,Nowak06,Nowak06a} and is also widely
applied to general evolutionary theories.

Error threshold\cite{Eigen77,Smith95} appears as an upper bound on
mutation rate, above which no effective selection can occur and the
quasispecies becomes unstable. It places limits on how genetic information
can be maintained and passed on from generations to generations and
restricts possible theories for the origins of life. However, it is
quite puzzling that some RNS viruses\cite{Smith95,Drake93,Dominigo96}
seem to have mutation rates close to the error threshold. The error
threshold for the Eigen model was first derived by ignoring mutations
of mutants back to the wild type, known as the assumption of no back
mutation\cite{Nowak06}. Later, the model was extensively studied
by mapping to equivalent statistical models\cite{Schuster85,Leuthausser86,Nowak89,Tarazona92,Baake97}
and the existence of the error threshold can be viewed as a phase
transition from the ordered phase (quasispecies) to the disordered
phase (no effective selection). Exact solution of the model\cite{Galluccio97,Saakian04,Saakian06}
shows that the error threshold is identical to that derived under
the assumption of no back mutation. In addition, the phase transition
is shown to be first-order\cite{Eigen00}.

Though the exact solution\cite{Galluccio97,Saakian04,Saakian06} is
available for the Eigen model, due to technical complications, it
is not yet clear why the error threshold derived within the assumption
of no back mutation is identical to the exact solution. Meanwhile,
it is desirable to devise a mean-field approach capturing the essence
of the phase transition so that the order parameter can be computed
(and understood) in the easiest way. Inspired by the fact that this
is a first-order phase transition, it is possible for one to come
up with a variational ansatz capturing the phase transition. In this
article, we make a full use of the symmetry and propose a simple variational
ansatz containing only two configurations in the sequence space. It
is rather remarkable that the variational approach captures the essence
of the phase transition including the error threshold and also the
evolution of the wild-type population. It is quite surprising that
error thresholds for different genome lengths, values of mutation
rate and relative fitness collapse onto a universal curve, giving
credit to the power of the variational ansatz. We also perform numerical
calculations to verify the results obtained by the variational ansatz
and find nice agreement between them.

This article is organized as follows. In Sec.~\ref{sec:Eigen-Model},
we briefly introduce the Eigen model and a gauge transformation to
an equivalent Hamiltonian with imaginary-time dynamics. In Sec.~\ref{sec:Variational-Ansatz},
the Hamiltonian can be solved by the proposed variation ansatz, giving
the error threshold and also the evolution of the wild-type population.
In Sec.~\ref{sec:No-Back-Mutation?}, we calculate the back-mutation
rate to the wild type and confirm that the error threshold remains
the same within the approximation of no back mutation. In Sec.~\ref{sec:Numerical-Verification},
we present a numerical approach to the Eigen model and compare the
results with the variational approach.

\section{Eigen Model\label{sec:Eigen-Model}}

Consider a DNA chain with genome length $L$. The primary structure
at each site of the chain takes on four different nucleotides ($G$,
$A$, $C$ and $U$). For simplicity, we choose to distinguish only
among purines ($R$) and pyrimidines ($Y$). Thus, the total size
of sequence space is reduced to $2^{L}$, yet still huge for numerical
simulations. A viral genome with typical length $L\sim1000$ will
lead to a (reduced) sequence space of the size $\sim10^{300}$. For
more complex forms of life, the size of the sequence space increases
tremendously and methods developed in statistical mechanics seem appropriate
and powerful to address the dynamics of the genome evolution.

The time-dependent relative population of a particular sequence $i$
is denoted as $x_{i}(t)$. The selection is on the genotype level
described by the fitness $f_{i}$ for each sequence. In general, the
fitness function can be rather complicated and evolves with time.
Here it is assumed to take on its time-averaged value and treated
as a constant. Eigen proposed a deterministic dynamics to describe
the relative populations of the sequences in the presence of mutations,
\begin{equation}
\frac{dx_{i}}{dt}=\sum_{j}\left(m_{ij}f_{j}-\phi\delta_{ij}\right)x_{j}\end{equation}
 where $x_{i}$ and $f_{i}$ are the relative population and the fitness
of the sequence $i$, while $\phi=\sum_{i}x_{i}f_{i}$ is the average
fitness for all sequences. The relative populations are normalized
with $\sum_{i}x_{i}=1$ initially and remain so under the time evolution
of the quasispecies equation. The mutation matrix $m_{ij}$ describes
the probability to mutate from sequence $j$ to sequence $i$ in a
generation and satisfies $\sum_{i}m_{ij}=1$. If only point mutations
are considered, it is straightforward to work out the mutation matrix
\begin{equation}
m_{ij}=u^{d_{ij}}(1-u)^{L-d_{ij}},\end{equation}
 where $u$ is the probability for a single point mutation to occur
within a generation and $d_{ij}$ is the Hamming distance\cite{Hamming80}
between the sequences. Since $d_{ij}=d_{ji}$, it is clear that the
mutation matrix $m_{ij}$ is also symmetric.

The quasispecies equation can be transformed into an imaginary-time
Schrödinger equation by the gauge transformation, \begin{equation}
\Psi_{i}(t)=\sqrt{f_{i}}x_{i}(t)e^{W(t)},\end{equation}
 where $\dot{W}(t)=\phi(t)$. Applying the gauge transformation to
the quasispecies equation, the dynamics of $\Psi_{i}(t)$ is captured
by an effective Hamiltonian $H_{ij}$, \begin{equation}
\frac{d\Psi_{i}}{dt}=-\sum_{j}H_{ij}\Psi_{j},\quad\mbox{with}\quad H_{ij}=-\sqrt{f_{i}f_{j}}m_{ij}.\end{equation}
 One notices that the above equation can be obtained from the usual
Schrödinger equation by replacing the time variable $t\to it$ (and
setting $\hbar=1$). Following the standard textbooks, the general
solution for the {}``imaginary-time'' Schrödinger equation is \begin{equation}
\Psi_{i}(t)=\sum_{n}c_{n}\Phi_{i}^{n}e^{-E_{n}t},\end{equation}
 where $E_{n}$ and $\Phi_{i}^{n}$ are the eigenvalues and eigenvectors
of the effective Hamiltonian $H_{ij}$. Note that the factor $\sqrt{f_{i}}$
in the gauge transformation is crucial to make the Hamiltonian symmetric.

Unlike the intrinsic oscillatory nature of the usual Schrödinger equation,
its imaginary-time version only keeps the ground state in the infinite-time
limit, \begin{equation}
\lim_{t\to\infty}\Psi_{i}(t)\:\:\longrightarrow\:\: c_{0}\Phi_{i}^{0}e^{-E_{0}t},\end{equation}
 and greatly simplifies the calculations. Making use of the normalization
condition, $\sum_{i}x_{i}=1$, the exponential factor in the gauge
transformation can be expressed as \begin{equation}
e^{W(t)}=\sum_{i}\frac{1}{\sqrt{f_{i}}}\Psi_{i}(t).\end{equation}
 The inverse gauge transformation can thus be found, \begin{eqnarray}
x_{i}(t) & = & \frac{1}{\sqrt{f_{i}}}\Psi_{i}(t)e^{-W(t)}\nonumber \\
 & = & \left(\frac{1}{\sum_{j}\Psi_{j}/\sqrt{f_{j}}}\right)\frac{1}{\sqrt{f_{i}}}\Psi_{i}(t).\end{eqnarray}
 In the infinite-time limit, only the ground-state wave function survives
and the relative populations are \begin{equation}
x_{i}^{*}=\lim_{t\to\infty}x_{i}(t)=\left(\frac{1}{\sum_{j}\Phi_{j}^{0}/\sqrt{f_{j}}}\right)\frac{1}{\sqrt{f_{i}}}\Phi_{i}^{0}.\end{equation}
 Note that the constant $c_{0}$ cancels itself out and does not appear
in $x_{i}^{*}$. To find the survival populations $x_{i}^{*}$, one
only needs to find the ground state $\Phi_{i}^{0}$ of the effective
Hamiltonian.

\section{Variational Ansatz\label{sec:Variational-Ansatz}}

In Eigen's original proposal, the wild-type sequence, say, $0$ has
maximum fitness $f_{0}=f_{M}$ and all other sequences $i\neq0$ have
a common background fitness $f_{i}=f<f_{M}$. Even though the single-peak
fitness landscape is simple, the size of the sequence space is still
huge $N_{s}=2^{L}$. Since the fitness landscape is obviously symmetric
around the wild-type peak, it is convenient to introduce the representation
for arbitrary configurations, \begin{equation}
\Psi=(\psi^{0},\psi^{1},...,\psi^{L}),\end{equation}
 where $\psi^{d}$ is a row vector representing states with Hamming
distance $d$ to the peak and thus has the dimension of $C_{d}^{L}=L!/d!(L-d)!$.

\begin{figure}
\includegraphics[width=7cm]{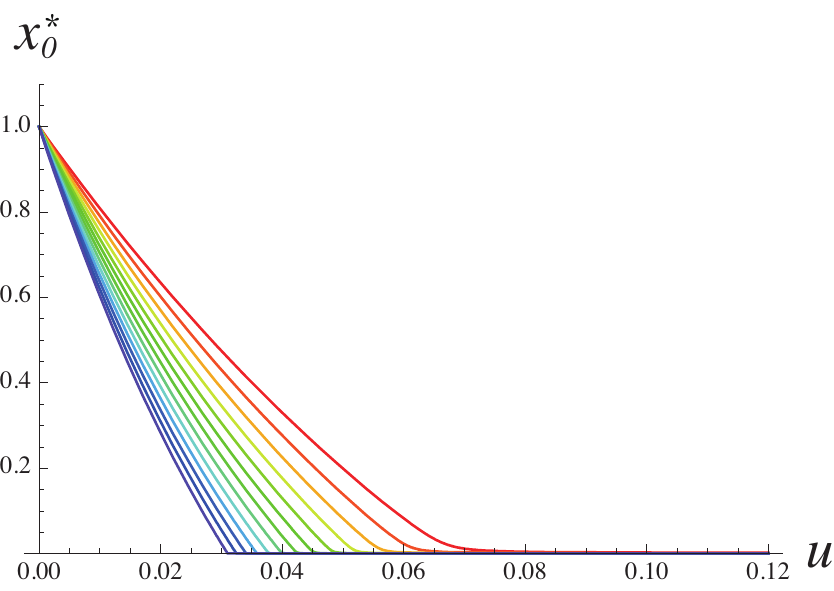} \includegraphics[width=7cm]{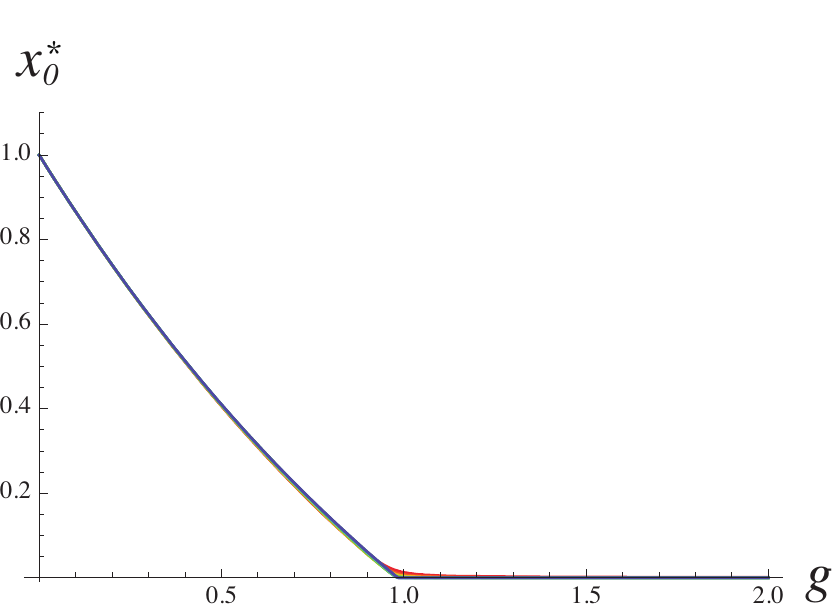}
\caption{The survival population $x_{0}^{*}$ of the wild-type sequence plotted
against the point mutation rate $u$ and the rescaled variable $g=uL/\ln r$.
The fitness landscape consists of a single sharp peak with relative
fitness $r=f_{M}/f=2$ to a uniform background fitness $f$ for all
other sequences. The genome length starts from $L=10$ (red) to $L=22$
(purple). It is clear that the crossover (from the localized quasispecies
to the extended state without selection) becomes sharper as the genome
length increases. For given relative fitness $r$, all curves for
$x_{0}^{*}(u,L)$ can be collapsed onto the universal scaling function
by changing the variable to $g=uL/\ln r$. It is clear that the mutation
threshold is $g_{c}=1$.}

\end{figure}

Since we are interested in the low-energy sector of the effective
Hamiltonian, it is sufficient to keep the so-called {}``$s$-wave''
sector\cite{Saakian06} only. That is to say, out of the many states
with the same Hamming distance to the wild-type sequence, one only
needs to keep the totally symmetric state for each Hamming distance,
\begin{eqnarray}
\Psi^{0s} & = & (\psi^{0},0,..,0)=(1,0,0,...,0),\\
\Psi^{1s}\!\! & = & \!\!(0,\psi^{1},0,...,0)=\!\!\sqrt{\frac{1}{L}}(0,1,1,..,1,0,0,..0),\end{eqnarray}
 and other $\Psi^{ns}$ can be constructed in a similar way. Writing
down the effective Hamiltonian explicitly, \begin{equation}
H_{ij}=-\sqrt{f_{i}f_{j}}u^{d_{ji}}(1-u)^{L-d_{ji}},\end{equation}
 one can construct the reduced $\left(L+1\right)\times\left(L+1\right)$
Hamiltonian in the $s$-wave sector \begin{equation}
H_{nm}=\sum_{ij}\Psi_{i}^{ns}H_{ij}\Psi_{j}^{ms},\end{equation}
 and numerically diagonalize the Hamiltonian to obtain the ground
state. The $s$-wave symmetry helps tremendously and reduces the size
of the relevant sequence space from $N_{s}=2^{L}$ to just $(L+1)$.
The comparison between the $s$-wave decomposition and the exact solution
will be discussed in Section \ref{sec:Numerical-Verification}.

Though the $s$-wave decomposition largely reduces the effective dimension
of the sequence space, it is still too complicated for analytic manipulations.
Here we choose two symmetric base functions, \begin{equation}
\Psi^{0}=(1,0,...,0),\qquad\Psi^{1}=\frac{1}{\sqrt{N_{s}-1}}(0,1,1,...,1),\end{equation}
 and use variational approach to estimate the error threshold for
the Eigen model. Construct the variational wave function for the ground
state \begin{equation}
\Psi_{i}=c_{0}\Psi_{i}^{0}+c_{1}\Psi_{i}^{1},\end{equation}
 so that the variational energy takes the form \begin{equation}
E(c_{1},c_{2},\lambda)=\sum_{n,m=0,1}c_{n}H_{nm}^{v}c_{m}+\lambda(c_{0}^{2}+c_{1}^{2}),\end{equation}
 where the Lagrange multiplier arises from the normalization constraint
$c_{0}^{2}+c_{1}^{2}=1$ and the variational Hamiltonian $H_{nm}^{v}=\sum_{ij}\Psi_{i}^{n}H_{ij}\Psi_{j}^{m}$
is a $2\times2$ real symmetric matrix, \begin{eqnarray}
H_{00}^{v} & = & -f_{M}(1-u)^{L},\\
H_{11}^{v} & = & -\frac{f}{N_{s}-1}\sum_{i,j\neq0}u^{d_{ij}}(1-u)^{L-d_{ij}}\nonumber \\
 & = & -\frac{f}{N_{s}-1}\left(\sum_{i\neq0,j}-\sum_{i\neq0,j=0}\right)u^{d_{ij}}(1-u)^{L-d_{ij}}\nonumber \\
 & = & -f+\frac{f}{N_{s}-1}\left[1-(1-u)^{L}\right],\\
H_{01}^{v} & = & H_{10}^{v}=-\sqrt{\frac{f_{M}f}{N_{s}-1}}\left[1-(1-u)^{L}\right].\end{eqnarray}
 Minimizing the variational energy $E(c_{1},c_{2},\lambda)$ is equivalent
to diagonalizing the $2\times2$ variational Hamiltonian $H_{nm}^{v}$,
which can be rewritten as \begin{equation}
H_{nm}^{v}=C\textbf{1}+f\left(\begin{array}{cc}
-\epsilon & -\Delta\\
-\Delta & \epsilon\end{array}\right),\end{equation}
 where the constant term does not affect the frequencies $x_{i}(t)$
after the inverse gauge transformation, \begin{equation}
C=-\frac{1}{2}\left[f_{M}m_{00}+f-\frac{f}{N_{s}-1}(1-m_{00})\right]\end{equation}
 with $m_{00}=(1-u)^{L}$ and will be dropped in the following calculations.
The dimensionless parameters $\epsilon$ and $\Delta$ in the variational
Hamiltonian are \begin{eqnarray}
\epsilon(u,L,r) & = & \frac{1}{2}\left[rm_{00}-1\right]+\frac{1-m_{00}}{2(N_{s}-1)},\\
\Delta(u,L,r) & = & \sqrt{\frac{r}{N_{s}-1}}\left[1-m_{00}\right],\end{eqnarray}
 where $r=f_{M}/f$ is the relative fitness of the dominant sequence
and $N_{s}=2^{L}$ is the size of the sequence space. The eigenvalues
of the variational Hamiltonian are \begin{equation}
E_{\pm}=\pm\sqrt{\epsilon^{2}+\Delta^{2}}.\end{equation}
 Therefore, the ground-state energy within the variational approximation
corresponds to the negative eigenvalue $E=E_{-}$ with the eigenvector
\begin{equation}
\frac{c_{0}}{c_{1}}=\frac{\Delta}{\sqrt{\epsilon^{2}+\Delta^{2}}-\epsilon}=\frac{\sqrt{\epsilon^{2}+\Delta^{2}}+\epsilon}{\Delta}.\end{equation}

With the coefficients $c_{0}$ and $c_{1}$ at hand, the ground state
within the variational approximation is \begin{equation}
\Phi^{v0}=\left(c_{0},\frac{c_{1}}{\sqrt{N_{s}-1}},\frac{c_{1}}{\sqrt{N_{s}-1}},...,\frac{c_{1}}{\sqrt{N_{s}-1}}\right).\end{equation}
 The relative population of the wild-type sequence in the infinite-time
limit can be computed from the variational ground state, \begin{eqnarray}
x_{0}^{*}(u,L,r) & = & \left(\frac{1}{\sum_{j}\Phi_{j}^{v0}/\sqrt{f_{j}}}\right)\frac{1}{\sqrt{f_{0}}}\Phi_{0}^{v0}\nonumber \\
 &  & \hspace{-15mm}=\frac{\sqrt{\epsilon^{2}+\Delta^{2}}+\epsilon}{\sqrt{\epsilon^{2}+\Delta^{2}}+\epsilon+\sqrt{r(2^{L}-1)}\Delta}.\end{eqnarray}
 As shown in Fig. 1, the survival population of the wild-type sequence
$x_{0}^{*}$ is plotted against the probability of point mutation
$u$ for different genome lengths ranging from $L=10$ to $L=22$
for the relative fitness $r\equiv f_{M}/f=2$.

The error threshold, where the quasispecies disappears, marks the
phase transition and, in theory, is only well defined in the thermal
dynamic limit with infinite genome length. Although there is no extra
numerical cost to choose a longer genome length $L$ within the variational
approach, we intentionally choose relatively short $L$s to highlight
the sharpness of the phase transition at finite genome lengths. Furthermore,
for chosen relative fitness $r$, the survival populations $x_{0}^{*}(u,L)$
for different $u$ and $L$ can be collapsed onto a universal curve
when plotted with the rescaled variable \begin{equation}
g=\frac{uL}{\ln r}.\end{equation}
 As is evident in Fig. 1, the collapse is almost perfect even for
such short genome lengths. The error threshold occurs at $g_{c}=1$,
\emph{i.e.,} $u_{c}L=\ln(f_{M}/f)$ which agrees with the exact solution.

\begin{figure}
\includegraphics[width=7cm]{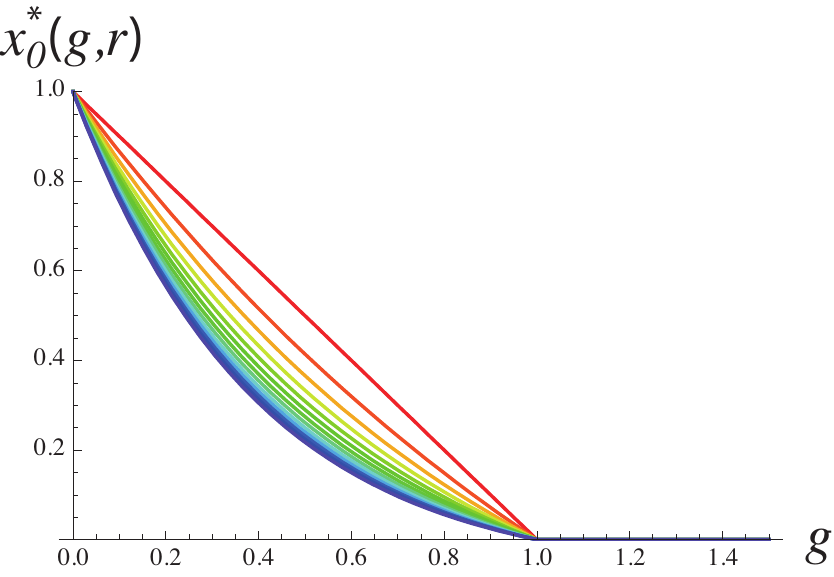} \caption{The universal scaling function $x_{0}^{*}(g,r)$ in the limit of infinite
genome length. The transition is sharp and the universal scaling function
varies with relative fitness from $r=1^{+}$ (red) to $r=13$ (purple).}

\label{infinite} 
\end{figure}

The universal curves for the survival population $x_{0}^{*}(g,r)$
can be derived by taking the genome length $L$ to infinity. Taking
$L\to\infty$ but keeping $g$ and $r$ finite, the stay-on probability
is $m_{00}=(1-u)^{L}\to r^{-g}$. Since the off-diagonal term $\Delta$
is negligibly small in this limit, the survival population of the
wild-type sequence is greatly simplified, \begin{eqnarray}
x_{0}^{*}(g,r) & = & \lim_{L\to\infty}x_{0}^{*}(u,r,L)\nonumber \\
 &  & \hspace{-15mm}=\frac{\frac{1}{2}\left[|r^{1-g}-1|+(r^{1-g}-1)\right]}{\frac{1}{2}\left[|r^{1-g}-1|+(r^{1-g}-1)\right]+r(1-r^{-g})}.\end{eqnarray}
 Since the relative fitness $r$ is greater than unity, the numerator
is zero for $g>1$ and the scaling function takes the simple form,
\begin{equation}
x_{0}^{*}(g,r)=\Theta(1-g)\:\frac{r^{1-g}-1}{r-1}.\end{equation}
 Magically, this form is identical to that derived under the assumption
of no back mutations to the wild-type sequence. The universal scaling
behavior within the variational approach is robust, hinting that the
simple variational ansatz captures the essence of the phase transition.

\begin{figure}
\includegraphics[width=7cm]{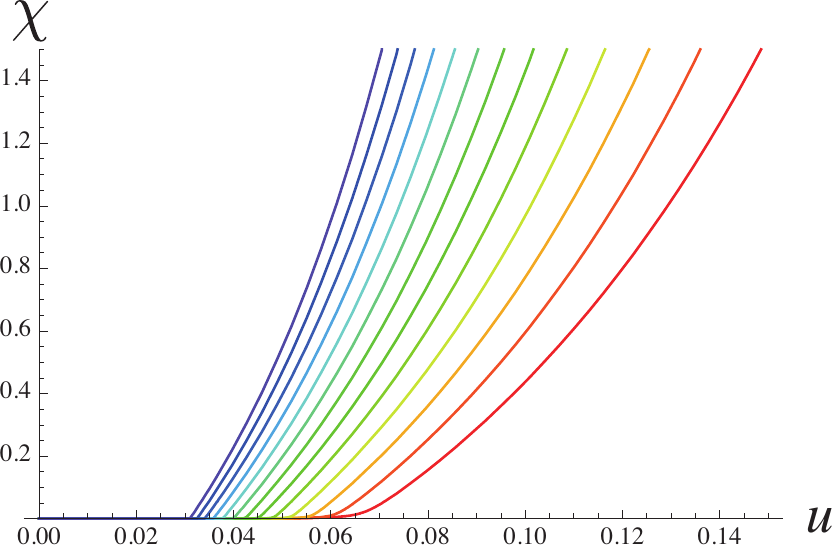} \includegraphics[width=7cm]{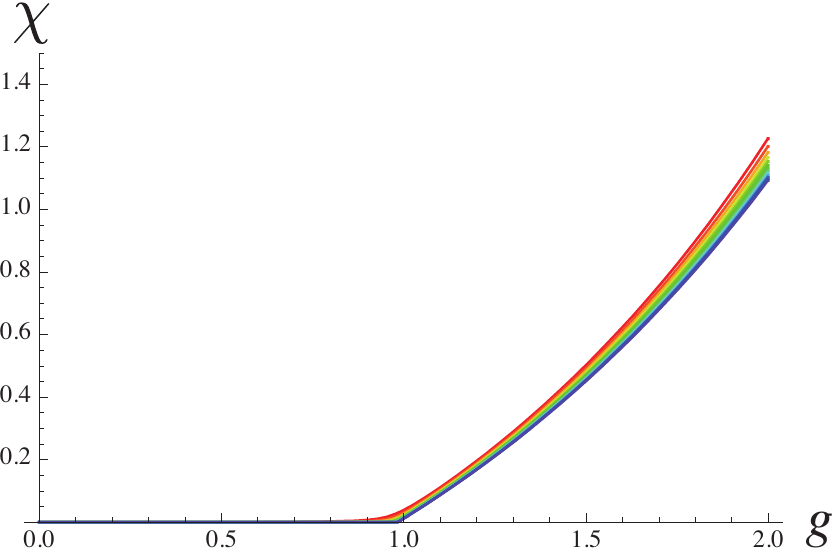}
\caption{The back-mutation ratio $\chi(u,L,r)$ plotted against the mutation
rate $u$ and the rescaled variable $g=uL/\ln r$. The fitness landscape
consists of a single sharp peak with relative fitness $r=f_{M}/f=2$
to a uniform background fitness $f$ for all other sequences. The
genome lengths range from $L=10$ (red) to $L=22$ (blue). For large
enough genome length, the back-mutation rate is indeed negligibly
small when compared with the rate to stay on the dominant sequence.
However, beyond the mutation threshold, the ratio $\chi$ grows exponentially
with the disappearance of quasispecies.}

\label{chi} 
\end{figure}

\section{No Back Mutation?\label{sec:No-Back-Mutation?}}

The variational ansatz also provides a helping hand to checking how
good the assumption of no back mutations is. A good indicator is the
ratio between the back-mutating rate and the rate to stay on the wild-type
sequence, \begin{equation}
\chi(u,L,r)=\frac{\sum_{j\neq0}m_{0j}f_{j}x_{j}^{*}}{m_{00}f_{0}x_{0}^{*}}.\end{equation}
 Within the variational approximation, all $x_{j}^{*}=x_{1}^{*}$
for $j\neq0$ and the summation in the numerator can be carried out
without difficulty, \begin{eqnarray}
\chi(u,L,r) & = & \frac{x_{1}^{*}f(1-m_{00})}{x_{0}^{*}f_{M}m_{00}}\nonumber \\
 & = & \sqrt{\frac{f}{f_{M}}}\frac{1-m_{00}}{m_{00}}\frac{c_{1}}{\sqrt{(N_{s}-1)}c_{0}}.\end{eqnarray}
 For $r=2$, the back-mutation indicator $\chi$ is shown in Fig.
3. When the quasispecies exists, the back-mutation rate is exponentially
small even for a finite genome length. Therefore, the no-back-mutation
assumption is justified in this regime. On the other hand, when the
mutation probability $u$ exceeds the error threshold, the indicator
$\chi$ grows exponentially and shows that the back-mutation rate
dominates the rate to stay on the wild-type sequence. The dynamics
is dominated by mutual mutations between different sequences and no
selection is at work. In this regime, the assumption of no back mutations
is inappropriate and the variational ansatz delivers a better description.
Again, when plotted against the rescaled variable $g=uL/\ln r$, it
is shown in Fig. 3 that all curves collapse onto a universal one with
the threshold $g_{c}=1$.

Similarly, by taking the genome length to infinity, the scaling function
for $\chi(g,r)$ can be derived. Note that the $L\to\infty$ limit
is taken while keeping both $g$ and $r$ finite. The only tricky
term in $\chi(g,r)$ is \begin{equation}
\lim_{L\to\infty}\frac{c_{1}}{\sqrt{(N_{s}-1)}c_{0}}=\Theta(g-1)\frac{1-r^{1-g}}{\sqrt{r}(1-r^{-g})}.\end{equation}
 Thus, the scaling function for the back-mutation indicator in the
infinite genome length is \begin{equation}
\chi(g,r)=\Theta(g-1)\:\left(r^{g-1}-1\right).\end{equation}
 The universal function $\chi(g,r)$ is zero in the presence of the
quasispecies, \emph{i.e.}, ignoring the back mutations to the wild-type
sequence is justified.

\section{Numerical Verification\label{sec:Numerical-Verification}}

\begin{figure}
\includegraphics[width=8cm]{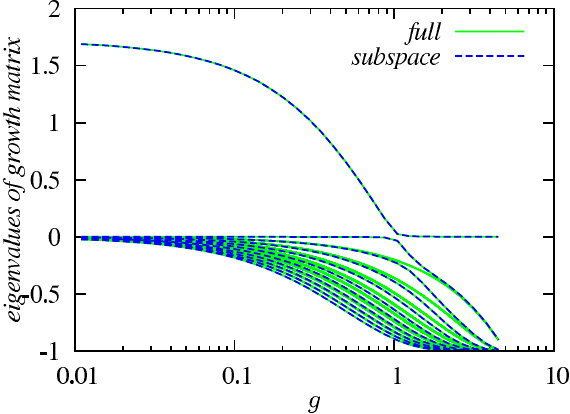} \caption{The eigenvalues of the $(L+1)\times(L+1)$ reduced matrix $H_{nm}$and
the $2^{L}\times2^{L}$ full growth matrix \textbf{G} ($G_{ij}\equiv m_{ij}f_{j}-1$)
at $L=10$ plotted against the scaled mutation parameter $g=uL/\ln r$.
We see an exact match of the eigenvalues of these two matrices and
some additional eigenvalues of the full matrix not matched by those
of the reduced one, that are likely corresponding to non-symmetric
modes of decay.}

\label{eigenvalue} 
\end{figure}

To check the validity of the variational ansatz, we also solve the
Eigen model numerically. We start with Eigen's proposal and study
the evolution of populations $X_{i}(t)$ (not relative populations)
in discrete generations, \begin{equation}
X_{i}(t+1)=\sum_{j}m_{ij}f_{j}X_{j}(t),\end{equation}
 where the mutation matrix $m_{ij}$ and the fitness function $f_{i}$
are defined previously. It is convenient to introduce the growth matrix
$\mathbf{G}$ defined as \begin{equation}
G_{ij}=m_{ij}f_{j}-\delta_{ij}.\end{equation}
 Positive eigenvalues of the growth matrix imply population growth
and negative ones mean diminishing populations. The stationary state
then corresponds to the eigenstate of $\mathbf{G}$ with the largest
eigenvalue.

To check the validity of the $s$-wave decomposition, we compute the
eigenvalues of $\mathbf{G}$ in the projected sequence space and compare
the results to the exact diagonalization in the full space. As shown
in Fig. 4, the $s$-wave decomposition works rather well. Near the
error threshold $g_{c}=1$, there are only two symmetric states coming
close to each other. A detail check reveals that these are just the
variational wave functions we chose before. Therefore, the numerical
results lend hands to the support of our variational ansatz. Although
the numerical results presented here are for evolution in discrete
time steps, one can prove (not shown here) that the stationary state
of the quasispecies equation gives identical configuration.

\begin{figure}
\includegraphics[width=7.5cm]{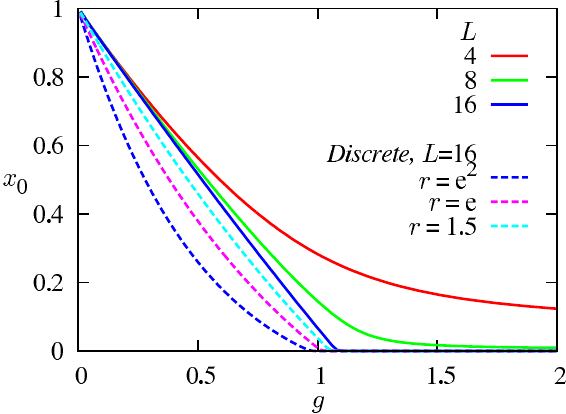} \caption{Survival population of the wild-type sequence $x_{0}$ plotted against
the rescaled variable $g$ in the Crow-Kimura model (solid lines)
and the discrete Eigen model (dashed lines). For genome length $L=16$,
the survival population from the discrete Eigen model is well described
by the universal scaling function in Fig. 2 and approaches that obtained
in the Crow-Kimura model when the relative fitness $r$ approaches
unity.}

\label{conti_species} 
\end{figure}

\begin{figure}
\includegraphics[width=7cm]{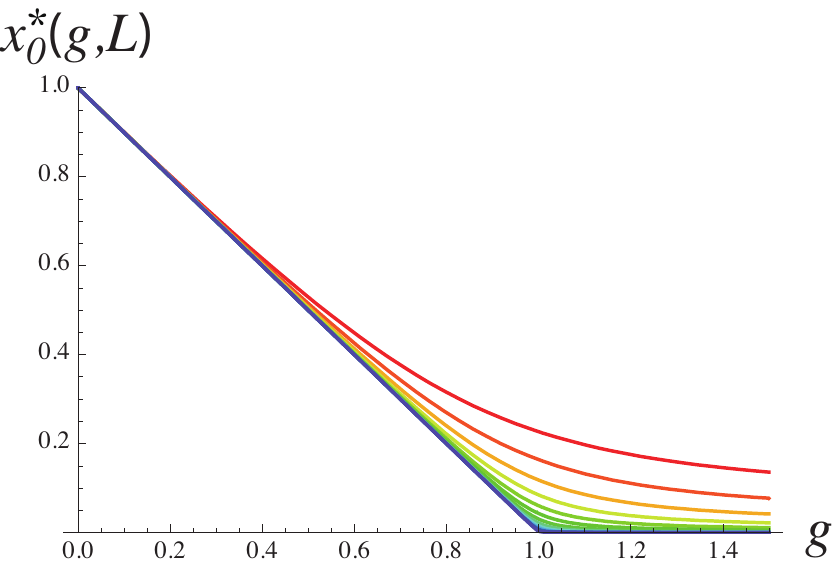} \caption{The survival population of the wild-type sequence $x_{0}$, varying
from $L=4$ (red) to $L=16$ (blue) in the continuous limit ($u\to0$
and $r\to1^{+}$ ). Note that the scaling function with infinite genome
length in the continuous limit is just the linear function $x_{0}^{*}(g)=1-g$
below the error threshold $g_{c}=1$.}

\label{conti} 
\end{figure}

It is interesting to go beyond the Eigen model and see whether the
variational ansatz still works. It is known in the literature, by
taking the evolutionary time step to the continuous limit, the discrete
Eigen model becomes the Crow-Kimura model, described by the set of
differential equations, \begin{equation}
\frac{dX_{i}}{dt}=G_{i}X_{i}+\sum_{j}\Gamma_{ij}X_{j},\end{equation}
 where $G_{i}=G+\delta_{i0}(G_{M}-G)$ is the grow rate for the $i$-th
sequence and $\Gamma_{ij}$ is the mutation-rate matrix with elements
$\Gamma_{ij}=\Gamma$ for $d_{ij}=1$, $\Gamma_{ii}=-\Gamma L$, and
zero otherwise. The rescaled variable $g$ is of crucial importance
in comparing results from different approaches. When taking the continuous
limit, the relations between $f_{i}$, $u$ (Eigen) and $G_{i}$,
$\Gamma$ (Crow-Kimura) are \begin{equation}
f_{i}=1+G_{i}\Delta t,\qquad u=\Gamma\Delta t,\end{equation}
 where $\Delta t$ is the microscopic time step before taking continuous
limit. The rescaled variable $g$ thus takes a slightly different
form, \begin{eqnarray}
g & = & \frac{uL}{\ln(f_{M}/f)}=\lim_{\Delta t\to0}\frac{\Gamma L\Delta t}{\ln(1+G_{M}\Delta t)-\ln(1+G\Delta t)}\nonumber \\
 & = & \frac{\Gamma L}{G_{M}-G}.\end{eqnarray}
 Plotted against the rescaled variable $g$, the survival populations
of the wild-type sequence for discrete Eigen model and the continuous
Crow-Kimura model are shown in Fig. 5. For discrete Eigen model with
genome length $L=16$, the survival population $x_{0}$ is very close
to the universal scaling function in Fig. 2. As the relative fitness
$r\equiv f_{M}/f$ approaches unity, the curve moves closer to that
obtained from the Crow-Kimura model. The results from the Crow-Kimura
model at shorter genome lengths are also plotted for later comparison.

To compare the stationary states in the quasispecies equation and
the Crow-Kimura model, we need to take $r\to1$ limit in the variational
ansatz, \begin{eqnarray}
\epsilon_{c}(g,L) & \equiv & \lim_{r\to1}\frac{\epsilon}{r-1}=\frac{1}{2}\left[1-g+\frac{g}{2^{L}-1}\right],\\
\Delta_{c}(g,L) & \equiv & \lim_{r\to1}\frac{\Delta}{r-1}=\frac{1}{\sqrt{2^{L}-1}}g.\end{eqnarray}
 After some algebra, the survival frequency of the dominant sequence
$x_{0}^{*}(g,L)$ in the continuous limit is \begin{equation}
x_{0}^{*}(g,L)=\frac{\sqrt{\epsilon_{c}^{2}+\Delta_{c}^{2}}+\epsilon_{c}}{\sqrt{\epsilon_{c}^{2}+\Delta_{c}^{2}}+\epsilon_{c}+g}.\end{equation}
 These curves for finite genome length $L$ are plotted in Fig. 6
and agree rather well with those obtained from Crow-Kimura model in
Fig. 5. Furthermore, the above expression becomes extremely simple
when the genome length goes to infinity, \begin{eqnarray}
\lim_{r\to1}x_{0}^{*}(g,r) & = & \Theta(1-g)\lim_{r\to1}\frac{r^{1-g}-1}{r-1}\nonumber \\
 & = & \Theta(1-g)\times(1-g).\end{eqnarray}
 It is quite remarkable that the universal scaling function with infinite
genome length in the continuous limit is just a straight line.

The numerical results from the discrete Eigen model and the Crow-Kimura
model agree very well with the variational ansatz we proposed for
the quasispecies equation. The success of the variational approach
relies on the fact that there are only two active states near the
phase transition (error threshold), verified by the exact diagonalization.
Although the exact solution for the Eigen model has been obtained
in the literature, our variational ansatz provides a simpler approach
at the cost of negligible errors. The no-back-mutation assumption
is examined and only remains justified in the presence of the quasispecies.
The universal scaling behavior near the error threshold was unexpected
and further confirmed the validity of the variational approach. Though
the emergence of the scaling behavior is not yet fully understood,
it is remarkable that the universal scaling function reigns even for
relatively short genome length such as $L=16$. Further investigations
may reveal the mechanism of the universal scaling and extract the
essential ingredients for the emergence of the quasispecies in molecular
evolution.

We acknowledge supports NSC-97-2112-M-007-022-MY3 from the National
Science Council in Taiwan. Financial supports and friendly environment
provided by the National Center for Theoretical Sciences in Taiwan
are also greatly appreciated.

\end{document}